\documentclass[a4paper,fleqn,usenatbib]{mnras}

\usepackage{newtxtext,newtxmath}

\usepackage[T1]{fontenc}
\usepackage{ae,aecompl}

\usepackage{graphicx}	
\usepackage{amsmath}	
\usepackage{amssymb}	

\title[Triggered SR and FRBs]{Triggered superradiance and fast radio bursts}

\author[M. Houde et al.]{
Martin Houde,$^{1}$\thanks{E-mail: mhoude2@uwo.ca}
Fereshteh Rajabi,$^{2,1}$
B. M. Gaensler,$^{3,4}$
Abhilash Mathews,$^{5}$
\newauthor and Victor Tranchant$^{6}$
\\
$^{1}$Department of Physics and Astronomy, The University of Western Ontario, 1151 Richmond Street, London, Ontario N6A 3K7, Canada\\
$^{2}$Institute for Quantum Computing, The University of Waterloo, 200 University Ave. West, Waterloo, Ontario N2L 3G1, Canada\\
$^{3}$Dunlap Institute for Astronomy and Astrophysics, University of Toronto, 50 St. George Street, Toronto, ON M5S 3H4, Canada\\
$^{4}$Department of Astronomy and Astrophysics, University of Toronto, Toronto, ON M5S 3H4, Canada\\
$^{5}$Plasma Science and Fusion Center, Massachusetts Institute of Technology, 77 Massachusetts Avenue, Cambridge, MA 02139, USA\\
$^{6}$\'Ecole Nationale Sup\'erieure d'\'Electrotechnique, d'\'Electronique, d'Informatique, d'Hydraulique et des T\'el\'ecommunications, 31071 Toulouse Cedex 7, France
}

\date{Accepted 2018 November 6. Received 2018 November 5; in original form 2018 August 19}

\pubyear{2018}

\begin{document}
\label{firstpage}
\pagerange{\pageref{firstpage}--\pageref{lastpage}}
\maketitle

\begin{abstract}
In this paper we develop a model for fast radio bursts (FRBs) based on triggered superradiance (SR) and apply it to previously published data of FRB 110220 and FRB 121102. We show how a young pulsar located at $\sim100$ pc or more from an SR/FRB system could initiate the onset of a powerful burst of radiation detectable over cosmological distances. Our models using the OH $^2\Pi_{3/2}$ ($J=3/2$) 1612 MHz and $^2\Pi_{3/2}$ ($J=5/2$) 6030 MHz spectral lines match the light curves well and suggest the entanglement of more than $10^{30}$ initially inverted molecules over lengths of approximately 300 au for a single SR sample. SR also accounts for the observed temporal narrowing of FRB pulses with increasing frequency for FRB 121102, and predicts a scaling of the FRB spectral bandwidth with the frequency of observation, which we found to be consistent with the existing data.  
\end{abstract}




\section{Introduction}\label{sec:introduction}

Fast Radio Bursts (FRBs) emanate from extragalactic sources and are characterized by short (from tens of $\mu$s to several ms) and powerful bursts of radiation detected at radio frequencies with bandwidths often spanning several hundreds of MHz \citep{Lorimer2007,Thornton2013,Spitler2014,Petroff2016,Ravi2016,Michilli2018,Gajjar2018}.  But despite increasing observational efforts aimed at their detection and the characterization of their host environments, the nature of FRBs remains elusive \citep{Fialkov2017a,Fialkov2017b}. 

Given the relatively low number of detected FRBs to date, most of our knowledge stems from studies of FRB 121102 \citep{Spitler2014}, the only such source that has so far been precisely located and had its host identified \citep{Chatterjee2017,Marcote2017,Tendulkar2017}. Importantly, FRB 121102 has been observed to repeat, though irregularly, and has yielded several detections at frequencies ranging from approximately 1 GHz to 8 GHz. Although at this time most questions remain unanswered, the wealth of data available for this source can already be used to put some constraints on potential FRB models. Accordingly, in this paper we expand on our previous application of Dicke's superradiance\footnote{Dicke's superradiance is a quantum mechanical effect that was introduced by R. H. Dicke in a seminal paper more than 60 years ago \citep{Dicke1954}. This phenomenon should not be confused with ``black hole superradiance,'' which has also been considered within the context of the FRB problem (e.g., \citealt{Rosa2018,Conlon2018}). See \citet{Brito2015} for a comprehensive treatment of black hole superradiance and its historical connection to Dicke's superradiance.} (SR) to the FRB problem \citep{Houde2018a}, while endeavouring to account for the new data. 

SR is a quantum phenomenon where a group of molecules (or atoms) become entangled through interactions with their common electromagnetic field \citep{Dicke1954,Dicke1964,Gross1982}. As a result molecules contained within a gas act as a unit, as opposed to independent entities, and behave cooperatively in a manner that strongly affects the characteristics of the gas and the radiation emanating from it. As was discussed in \citet{Houde2018a} (see also \citealt{Rajabi2016A,Rajabi2016B,Rajabi2017}), SR will take place when the following conditions are met: \emph {i)} the population levels of the relevant molecular transition must be in a state of inversion, \emph {ii)} there must be sufficient velocity coherence between the molecules to allow for their interaction and \emph {iii)} the time-scales for relaxation (e.g., collisions) and dephasing (e.g., Doppler broadening) processes must be longer than that characterizing SR. Because of these requirements, it follows that astronomical evidence for SR are more likely to be, and have been, detected in regions harbouring astronomical masers \citep{Rajabi2016B,Rajabi2017}. Under such circumstances SR will mainly manifest itself through short and powerful bursts of radiation, reminiscent of those characterizing FRBs. 

SR bursts emanating from a gas containing $N$ molecules have a temporal duration that scales as $\propto \tau_\mathrm{sp}/N$, with $\tau_\mathrm{sp}$ the spontaneous emission time-scale of the molecular transition under consideration, and an intensity $\propto N^2 I_0$, with $I_0$ the intensity resulting from the spontaneous emission of a single molecule. For these reasons, SR can be described as a coherent and cooperative spontaneous emission of the gas acting as a whole. This is in contrast to the independent spontaneous emission from the individual molecules, which would take place over a time-scale $\tau_\mathrm{sp}$ with an intensity $N I_0$. 

In \citet{Houde2018a} we successfully modelled FRB 110220 \citep{Thornton2013} and FRB 150418 \citep{Keane2016} using an SR model based on the OH $^2\Pi_{3/2}$ ($J=3/2$) 1612 MHz spectral line. We showed that coherence can build up in OH samples when a critical column density of inverted population is exceeded, i.e., $nL \geq \left(nL\right)_{\mathrm{crit}}$. Here, we consider an alternative scenario where SR can ensue even when $nL < \left(nL\right)_{\mathrm{crit}}$ if coherence is injected into the SR sample through an external coherent pulse. This pulse acts as a trigger by polarizing the system and lowering the critical column density of the inverted population, as will be shown in Secs. \ref{sec:methodology}  and \ref{sec:results}. 

Triggered SR has been extensively investigated in laboratory experiments \citep{Carlson1980, Benedict1996, Gross1982} and has been shown to be an effective way of inducing strong bursts of coherent radiation. Here, we extend this formalism to astronomical settings and FRBs. In what follows, we describe our methodology in Sec. \ref{sec:methodology}, followed by a discussion of our results in Sec. \ref{sec:results}. In the latter we apply our triggered SR model to existing data of FRB 110220 and FRB 121102, discuss the characteristics of the source of the triggering pulse, and also demonstrate how SR can account for other characteristics of FRBs (e.g., polarization levels, and temporal duration and spectral bandwidth as a function of frequency). Finally, we end with a summary and conclusion in Sec. \ref{sec:conclusion}.  

\section{Methodology}\label{sec:methodology}

The evolution of a one-dimensional SR large sample can be calculated using the following equations for (half of) the population inversion density, polarization and electric field as a function of position $z$ and retarded time $\tau=t-z/c$ 
\begin{eqnarray}
     \frac{\partial\hat{\mathbb{N}}}{\partial\tau} & = & \frac{i}{\hbar}\left(\hat{P}_0^+\hat{E}_0^+-\hat{E}_0^-\hat{P}_0^-\right)-\frac{\hat{\mathbb{N}}}{T_1}+\hat{\Lambda}_\mathbb{N} \label{eq:dN/dt} \\     \frac{\partial\hat{P}_0^+}{\partial\tau} & = & \frac{2id^2}{\hbar}\hat{E}_0^-\hat{\mathbb{N}}-\frac{\hat{P}_0^+}{T_2} \label{eq:dP/dt} \\
    \frac{\partial\hat{E}_0^+}{\partial z} & = & \frac{i\omega_0}{2\epsilon_0c}\hat{P}_0^- \label{eq:dE/dt}
\end{eqnarray}

\noindent obtained using the Heisenberg representation \citep{Arecchi1970,Gross1982,Benedict1996,Rajabi2016B,Houde2018a}. In the derivation of equations (\ref{eq:dN/dt})-(\ref{eq:dE/dt}), the so-called Maxwell-Bloch equations, the slowly-varying-envelope-approximation (SVEA) was used, while the polarization and electric field were modelled, at resonance, with
\begin{eqnarray}
     \mathbf{\hat{P}}^\pm\left(z,\tau\right) & = & \hat{P}_0^\pm\left(z,\tau\right)e^{\pm i\omega_0\tau}\boldsymbol{\varepsilon}_d \label{eq:P_svea} \\
    \mathbf{\hat{E}}^\pm\left(z,\tau\right) & = & \hat{E}_0^\pm\left(z,\tau\right)e^{\mp i\omega_0\tau}\boldsymbol{\varepsilon}_d, \label{eq:E_svea}
\end{eqnarray}

\noindent where $\boldsymbol{\varepsilon}_d=\mathbf{d}/\left|d\right|$ is the unit polarization vector characterizing the underlying molecular transition of frequency $\omega_0$ and electric dipole moment $d$. The timescales $T_1$ and $T_2$ are for losses in the inverted population and polarization, respectively, through non-coherent processes (e.g., collisions and Doppler broadening), while the non-coherent pumping rate of the population inversion density is accounted for with $\hat{\Lambda}_\mathbb{N}$. For all the calculations and models presented in this paper $\hat{\Lambda}_\mathbb{N}$ was set to a constant value.

In the calculations presented in this paper the population inversion density is initially assumed $0$ across the sample and eventually settles to a steady-state value $\hat{\mathbb{N}}_0\simeq\hat{\Lambda}_\mathbb{N}T_1$ (at $\tau=0$ in our numerical computations). At that time the polarization is $Nd/\left(2V\right)\sin\left(2/\sqrt{N}\right)$, with $N$ the number of molecules in the inverted population and $V$ the volume, while the electric field is set to $0$ across the sample. A time-varying triggering pulse $\hat{E}_0^+\left(z=0,\tau\right)$ is then applied at the entrance of the SR sample (i.e., at $z=0$) at some later time. 

We arbitrarily chose the following function for the triggering pulse
\begin{equation}
    \hat{E}_0^+\left(z=0,\tau\right) = i\hat{E}_0\cosh^{-2}\left[\left(\tau-\tau_0\right)/T_\mathrm{t}\right],\label{eq:trigger}
\end{equation}

\noindent where $\hat{E}_0$, $\tau_0$ and $T_\mathrm{t}$ are the triggering pulse's amplitude, the retarded time at its centre (i.e., at its peak) and its duration parameter, respectively. Although the strength and width of the triggering pulse have an impact in the response of the SR sample, its exact shape has little effect for the cases considered in this paper. It is also important to note that the triggering pulse is propagating longitudinally in the SR sample along the $z$-axis.

Given equations (\ref{eq:dN/dt})-(\ref{eq:dE/dt}), the aforementioned initial conditions for $\hat{\mathbb{N}}$, $\hat{P}_0^+$ and $\hat{E}_0^+$, as well as the amplitude of the inversion pump $\hat{\Lambda}_\mathbb{N}$ and equation (\ref{eq:trigger}) for the triggering pulse, we computed the evolution of an SR sample of length $L$ up to a retarded time $\tau_\mathrm{max}$ using established numerical integration methods. That is, we transformed the three evolution equations into a system of ordinary differential equations, built an evenly spaced Cartesian grid for the spatial  $0\leq z\leq L$ and retarded time $0\leq\tau\leq\tau_\mathrm{max}$ coordinates, and used a fourth-order Runge-Kutta method to solve the system of equations. The computations were performed by moving forward in $\tau$ for a particular discretized position $z_k$ ($k$ is an integer), where both the population inversion density and the polarization were evaluated at the next grid point in $\tau$, and the electric field at the next spatial point $z_{k+1}$. This continued until $\tau=\tau_\mathrm{max}$ was reached, at which point the process was repeated for $z=z_{k+1}$ and $\tau=0$, and so on until the entire grid was covered \citep{Mathews2017}.

This grid discretization method analogous to the method of lines implies that backward propagation is negligible \citep{Haroche1977}, and is therefore well adapted to SR in a mirrorless astronomical setting with a triggering pulse propagating in the forward direction.

As was the case in our previous applications of SR to astrophysics \citep{Rajabi2016A,Rajabi2016B,Rajabi2017,Rajabi2016Thesis,Houde2018a}, our SR samples are assumed to have a cylindrical geometry of length $L\gg\lambda$, with $\lambda$ the wavelength of radiation. The radius of a sample is constrained by imposing a Fresnel number of unity (i.e., the cross section of the cylinder is given by $A=\lambda L$), a necessary condition for preserving phase coherence along the length of the sample \citep{Gross1982,Rajabi2017}. 

The models and fits to the data presented in the next section were performed by adjusting the steady-state level of the inverted population density prior to the arrival of the triggering pulse (i.e., $\hat{\mathbb{N}}_0\simeq\hat{\Lambda}_\mathbb{N}T_1$), the length of the SR sample $L$, the dephasing time-scale $T_2$, and the parameters of the triggering pulse (i.e., $\hat{E}_0$ and $T_\mathrm{t}$ in equation (\ref{eq:trigger})). Finally, the models were scaled in intensity to the corresponding data.

\section{Results and Discussion}\label{sec:results}

In previous works on SR, whether dealing with star-forming regions \citep{Rajabi2017}, the surroundings of evolved stars \citep{Rajabi2016B} or FRBs \citep{Houde2018a}, we consistently focused on situations where SR is initiated when the inversion level is high enough to bring the column density $nL$ of the inverted population ($n=N/\left(AL\right)$) above the critical threshold $\left(nL\right)_\mathrm{crit}$ where the time-scale of SR
\begin{equation}
    T_\mathrm{R}=\tau_\mathrm{sp}\frac{8\pi}{3nL\lambda^2}  \label{eq:TR}
\end{equation}

\noindent becomes small enough in comparison to the non-coherent relaxation/dephasing time-scales $T_1$ and $T_2$. As before, in equation (\ref{eq:TR}) $\tau_\mathrm{sp}$ is the time-scale of spontaneous emission for the underlying molecular transition (i.e., $\tau_\mathrm{sp}$ is the inverse of the Einstein coefficient of spontaneous emission). 

The initiation of SR could thus be caused, for example, by a sufficiently strong increase in the pumping rate (i.e., $\hat{\Lambda}_\mathbb{N}$) over a time-scale $T_\mathrm{p}$ approximately matching the duration of the SR burst that ensues (or shorter). As previously mentioned this burst is characterized by a powerful coherent radiation of intensity proportional to $N^2$, and is emitted over a beam of small angular extent. In contrast, a similar region, i.e., one harbouring an inverted population with sufficient velocity coherence, for which $nL<\left(nL\right)_\mathrm{crit}$ will not see the onset of an SR burst but will instead be host to a steady-state astronomical maser with non-coherent intensity proportional to $N$.

However, an increase in the inverted column density is not the only way SR can be initiated. That is, even when $nL<\left(nL\right)_\mathrm{crit}$ it is possible to bring conditions favourable to the onset of SR by ``injecting'' coherence into the sample. To better understand how this comes about, let us consider the special case when we set $T_1=T_2\equiv T^\prime$ and $\hat{\Lambda}_\mathbb{N}=0$ in equations (\ref{eq:dN/dt})-(\ref{eq:dE/dt}), with the initial population inversion density $\hat{\mathbb{N}}\left(z,0\right)=n/2$. It can then be shown that 
\begin{equation}
    \left(nL\right)_\mathrm{crit}\approx \frac{2\pi}{3\lambda^2}\frac{\tau_\mathrm{sp}}{T^\prime}\left|\ln\left(\frac{\theta_0}{2\pi}\right)\right|^2 \label{eq:nL_crit}
\end{equation}

\noindent with $\theta_0$ the so-called initial Bloch angle. It is common to set $\theta_0=2/\sqrt{N}$ to account for initial polarization fluctuations due to spontaneous emission within the sample at $\tau=0$ \citep{Gross1976,Gross1982}. Since the number of molecules is large (e.g., of the order of at least $10^{30}$ for FRBs; see below), $\theta_0$ is a very small number implying an initial polarization
\begin{equation}
    \hat{P}_0^+\left(z,0\right) \simeq \frac{Nd}{2V}\theta_0;\label{eq:P0_initial}
\end{equation}

\noindent see Appendix A of \citet{Houde2018a} for more details. 

It becomes clear from equations (\ref{eq:nL_crit}) and (\ref{eq:P0_initial}) that inducing a larger polarization at $\tau=0$ will effectively reduce the critical inverted column density level and facilitate the initiation of SR. This is not surprising since a higher polarization implies a larger total electric dipole moment (per unit volume) in the sample, which results from having a larger number of individual molecular electric dipoles oscillating in a well-defined phase relationship. In other words, a higher polarization level implies increased coherence within the SR sample.

Still using the special conditions where $T_1=T_2\equiv T^\prime$, it can also be shown that the electric field and the Bloch angle are simply related through (see equation (A33) in \citealt{Houde2018a})
\begin{equation}
    \theta\left(z,\tau\right) = -\frac{i2d}{\hbar}\int_{-\infty}^{\tau}\hat{E}_0^+\left(z,\tau^\prime\right)d\tau^\prime.\label{eq:theta}
\end{equation}

\noindent It therefore follows that the presence of a triggering pulse of the type given in equation (\ref{eq:trigger}) at the entrance of the SR sample (i.e., at $z=0$) will set the value of the initial Bloch angle to
\begin{equation}
    \theta_0 = \frac{4d}{\hbar}\hat{E}_0T_\mathrm{t}\label{eq:theta0-trigger}
\end{equation}

\noindent and lower $ \left(nL\right)_\mathrm{crit}$ enough to allow for the emission of an SR radiation burst by appropriately choosing $\hat{E}_0T_\mathrm{t}$.

We now apply this triggered SR formalism, in the more general case where $T_1\neq T_2$, to previously published data of FRB 110220 and FRB 121102. 

\subsection{Applications to FRB data}

\subsubsection{FRB 110220}\label{sec:FRB110220}

We first develop a triggered SR model to the previously published data for FRB 110220 detected at $\sim 1.4\,\mathrm{GHz}$ \citep{Thornton2013}. As was done in \citet{Houde2018a}, we select the OH $^2\Pi_{3/2}$ ($J=3/2$) 1612 MHz spectral line for our calculations, while keeping in mind that any other lines in the corresponding frequency range known (or to be shown) to exhibit population inversion could also be used for the analysis.

We need to set a few parameters that come in the model. First, damping in the SR signal is due to the finite values associated with the relaxation $T_1$ and dephasing $T_2$ time-scales. Since losses in the inverted population (i.e., relaxation) are likely due to collisions, it is to be expected that $T_1$ will be much greater than the duration of the intensity burst ($\sim10\,\mathrm{ms}$). Evidently, $T_1$ will vary as a function of the (unknown) gas density and temperature of the medium hosting the FRB system. But the fact that $T_1\gg T_\mathrm{R}$ implies that this time-scale will have practically no effect on the shape of the intensity curve produced by our SR model. We therefore arbitrarily set $T_1=10\,\mathrm{s}$ (i.e., corresponding to a density of $\sim10^9\,\mathrm{cm}^{-3}$ for a relative velocity of $\sim1\,\mathrm{km\,s}^{-1}$ between colliding partners), although significantly different values (e.g., $T_1=1$ or $100\,\mathrm{s}$) would produce the same outcome. Secondly, we must also set an appropriate level for the population inversion density prior to the arrival of the triggering pulse. To do so, we choose an inversion density of $\sim 1\,\mathrm{cm}^{-3}$ for a molecular population spanning a velocity range of approximately $1\,\mathrm{km\,s}^{-1}$ (or $\sim5000\,\mathrm{Hz}$ at 1612 MHz). For a spectral width of $\sim50\,\mathrm{Hz}$, which approximately matches the temporal profile of FRB 110220 (and therefore that from a single SR sample; see below), this corresponds to a population inversion density of approximately $0.01\,\mathrm{cm}^{-3}$, which we adopt for our calculations. We here again emphasize that a wide range of values are equally applicable for our SR model since the time-scale of SR (i.e., $T_\mathrm{R}$) is set by the column density of the inverted population $nL$. That is, for a given $T_\mathrm{R}$ a change in $n$ can be accommodated by the opposite change in $L$.

\begin{center}
   \begin{figure}
        \includegraphics[width=\columnwidth]{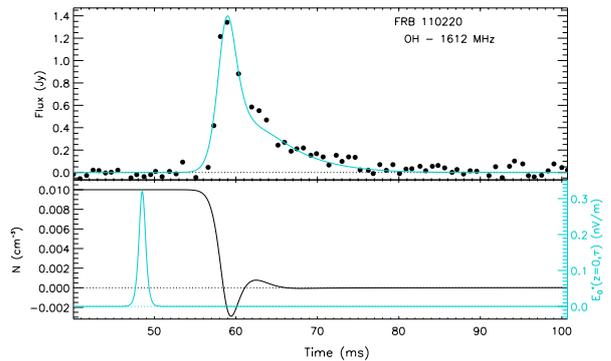}
        \caption{Triggered SR model for FRB 110220 \citep{Thornton2013}. Top: The black dots and cyan solid curve trace, respectively, the data and the resulting fit. Bottom: The black and cyan curves, respectively, show the inversion level and the triggering pulse, with corresponding scales on the left and right vertical axes. The model parameters are $L=4.2\times10^{15}$ cm ($280$ au), $T_1=10$ s and $T_2=1.2$ ms. The triggering pulse has an amplitude of $0.32\,\mathrm{nV}\,\mathrm{m}^{-1}$ and a width $T_\mathrm{t}=0.54$ ms. The inversion level prior to the trigger corresponds to approximately $1$ cm$^{-3}$ for a population spanning $1\,\mathrm{km}\,\mathrm{s}^{-1}$. The column density of the inverted population is $4.2\times10^{13}\,\mathrm{cm}^{-2}$. The model is scaled to the data.}
        \label{fig:FRB110220}
    \end{figure}
\end{center}

Figure \ref{fig:FRB110220} shows the result of our analysis using the model described in Sec. \ref{sec:methodology} to fit the data of \citet{Thornton2013} for FRB 110220. While initially unable to sustain an SR event, we find that an SR sample of length $L=4.2\times10^{15}$ cm ($280$ au) and inverted column density $nL = 4.2\times10^{13}\,\mathrm{cm}^{-2}$ subjected to a triggering pulse of $0.32\,\mathrm{nV}\,\mathrm{m}^{-1}$ in amplitude and $T_\mathrm{t}=0.54$ ms in width (see equation (\ref{eq:trigger})) will emit a strong burst of radiation approximately 10 ms after the arrival of the trigger (still using the retarded time $\tau$). Such a delay before the appearance of the radiation burst is a signature of SR, a phenomenon which does not take place for non-coherent radiation (e.g., astronomical masers). The dephasing time-scale $T_2=1.2$ ms is most likely set by Doppler motions within the sample \citep{Houde2018a}. The resulting SR model (solid cyan curve) matches the data (black dots) well, as can be seen in the top panel of the figure. The black and cyan curves in the bottom panel, respectively, show the inversion level at the end-fire of the sample (i.e., at $z=L$) and the triggering pulse at its entrance (i.e., at $z=0$), with corresponding scales on the left and right vertical axes.

It is important to note from Figure \ref{fig:FRB110220} that the symmetric shape of the triggering pulse is not transferred to the SR burst, which displays a significant asymmetry in its profile. Although the amplitude of the trigger is ``amplified'' by a factor of $\sim10^8$ from the entrance to the end-fire of the SR sample, the shape of the burst is largely a result of the characteristic response of the system to the excitation.

The cross-section of our model SR sample is small with a radius of 1580 km, while it contains on the order of $10^{30}$ entangled molecules that cooperatively emit a pulse of peak integrated intensity a few times $10^{-30}\,\mathrm{W\,m}^{-2}$ from a fiducial distance of 1 Gpc. As was discussed in \citet{Houde2018a}, regions hosting inverted populations responsible for astronomical (mega)masers are known to be much larger (several au). Consequently, the arrival of the triggering pulse and the ensuing initiation of SR will cause the aforementioned region to break into a very large number of simultaneously triggered SR samples \citep{Houde2018b}. We thus expect that the solid angle spanned by the radiation emitted from a region hosting the multiple SR samples will be defined by its geometry, in a manner similar to astronomical masers. This further implies that our SR model easily matches the detected flux density for FRB 110220. The same conclusion was reached by \citet{Houde2018a}, who also fitted this source with an SR model where the initial population inversion density was such that $nL>\left(nL\right)_\mathrm{crit}$. This earlier model took advantage of the sine-Gordon solution to the Maxwell-Bloch system of equations (i.e., when $T_1=T_2\equiv T^\prime$ and $\hat{\Lambda}_\mathbb{N}=0$ in equations (\ref{eq:dN/dt})-(\ref{eq:dE/dt})) and, although it does not provide as much freedom in the selection of the relaxation/dephasing parameters, also provided a good fit to the data. 

\subsubsection{FRB 121102}\label{sec:FRB121102}

As mentioned in Sec. \ref{sec:introduction}, FRB 121102 is, so far, the only source for which a precise location and redshift ($z=0.193$, \citealt{Tendulkar2017}) are known, and for which a large number of bursts have been detected in several frequency bands \citep{Spitler2014,Scholz2016,Law2017,Michilli2018,Gajjar2018}. The significant amount of available data for this object represents the best source for testing the many theoretical models put forth to explain the origin of FRBs. Accordingly, we now apply the triggered SR formalism to the Discovery Burst of FRB 121102 detected at $\sim1.4\,\mathrm{GHz}$ \citep{Spitler2014} and to recent observations from \citet{Michilli2018} obtained at 4--5 GHz. 

\begin{center}
   \begin{figure}
        \includegraphics[width=\columnwidth]{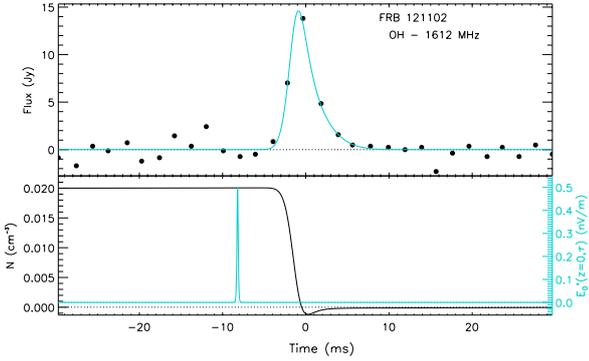}
        \caption{Same as Fig. \ref{fig:FRB110220} but for the Discovery Burst of FRB 121102 \citep{Spitler2014}. The model parameters are $L=4.45\times10^{15}$ cm ($\sim300$ au), $T_1=10$ s and $T_2=0.5$ ms. The triggering pulse has an amplitude of $0.5\,\mathrm{nV}\,\mathrm{m}^{-1}$ and a width $T_\mathrm{t}=0.1$ ms. The inversion level prior to the trigger corresponds to approximately $1$ cm$^{-3}$ for a population spanning $1\,\mathrm{km}\,\mathrm{s}^{-1}$. The column density of the inverted population is $8.9\times10^{13}\,\mathrm{cm}^{-2}$.}
        \label{fig:FRB121102-OH1612}
    \end{figure}
\end{center}

We once again choose the OH $^2\Pi_{3/2}$ ($J=3/2$) 1612 MHz transition for the Discovery Burst, although, as was the case for FRB 110220, other lines would do equally well. We kept the same values as before for the relaxation time-scale (i.e., $T_1=10\,\mathrm{s}$) and molecular population inversion density (i.e., corresponding to $1\,\mathrm{cm}^{-3}$ within $1\,\mathrm{km\,s}^{-1}$; see Sec. \ref{sec:FRB110220}). But the shorter duration of this burst ($\sim5\,\mathrm{ms}$) in comparison to FRB 110220 implies a larger spectral width of $\sim100\,\mathrm{Hz}$ for a single SR sample and, therefore, an increased value of $0.02\,\mathrm{cm}^{-3}$ for the population inversion density in that bandwidth prior to the arrival of the triggering pulse. Although a similar trigger as for FRB 110220 would also work, we opted for a narrower triggering pulse of width $T_\mathrm{t}=0.1$ ms (see below) and amplitude $0.5\,\mathrm{nV}\,\mathrm{m}^{-1}$ to initiate the SR burst. Otherwise, the other parameters needed for the fit were similar to those for the previous FRB 110220 model, except for the dephasing time-scale $T_2$. That is, our new model has $T_2=0.5\,\mathrm{ms}$, while the length of the sample was set to $L=4.45\times10^{15}\,\mathrm{cm}$ (or $\sim300$ au). The resulting fit is shown in Figure \ref{fig:FRB121102-OH1612}. 

We note that, in comparison to FRB 110220, the higher level of population inversion, and corresponding inverted column density $nL=8.9\times10^{13}\,\mathrm{cm}^{-2}$, resulted in a faster response (i.e., narrower pulse) and a shorter delay of approximately 7 ms before the appearance of the SR burst after the arrival of the trigger (in retarded time). Finally, the cross section of the SR sample has a radius of 1620 km, while $\sim8\times10^{30}$ entangled molecules are partaking in the emission process. 

\begin{center}
   \begin{figure}
        \includegraphics[width=\columnwidth]{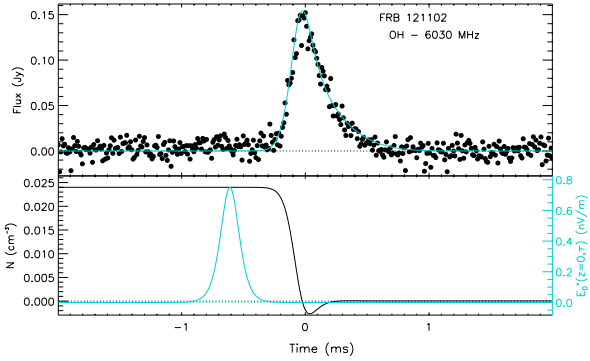}
        \caption{Same as Fig. \ref{fig:FRB121102-OH1612} but for FRB 121102 Burst 3 from \citet{Michilli2018}. The model parameters are $L=4.45\times10^{15}$ cm ($\sim300$ au), $T_1=10$ s and $T_2=0.05$ ms. The triggering pulse has an amplitude of $0.75\,\mathrm{nV}\,\mathrm{m}^{-1}$ and a width $T_\mathrm{t}=0.1$ ms. The inversion level prior to the trigger corresponds to approximately $0.12$ cm$^{-3}$ for a population spanning $1\,\mathrm{km}\,\mathrm{s}^{-1}$. The column density of the inverted population is $1.1\times10^{14}\,\mathrm{cm}^{-2}$.}
        \label{fig:FRB121102-OH6030}
    \end{figure}
\end{center}

In Figure \ref{fig:FRB121102-OH6030} we adapt our SR model to Burst 3 found in Extended Data Figure 1 of \citet{Michilli2018}, which resulted from observations of FRB 121102 at 4--5 GHz. Given the spectral band of observation we must choose a transition that scales correspondingly with frequency. We therefore selected the OH $^2\Pi_{3/2}$ ($J=5/2$) 6030 MHz spectral line commonly used in maser studies for our analysis \citep{Gray2012}. \footnote{Given the redshift associated to FRB 121102 the OH $^2\Pi_{3/2}$ ($J=5/2$) 6030 MHz spectral line would end up on the upper end of the spectral band under consideration in the observer's rest frame. This is consistent with the assumption that regions hosting SR/FRB systems must contain gas exhibiting motions over a wide velocity range \citep{Houde2018a}. We note, however, that the group of OH $^2\Pi_{1/2}$ ($J=7/2$) lines in the vibrational ground state at about 5.5 GHz would end up at lower frequencies (i.e., $\sim4.6$ GHz).} Since the SR characteristic time-scale varies with the frequency $\nu$ and the time-scale of spontaneous emission $\tau_\mathrm{sp}$ as $T_\mathrm{R}\propto \tau_\mathrm{sp}\nu^2$, we expect the OH 6030 MHz transition to be almost 10 times ``faster'' than the OH 1612 MHz line for a given inverted column density $nL$ (see eq. \ref{eq:TR}; the Einstein coefficients for spontaneous emission for the OH 1612 MHz and OH 6030 MHz transitions are $1.282\times10^{-11}\, \mathrm{s}^{-1}$ and $1.524\times10^{-9}\, \mathrm{s}^{-1}$, respectively). In other words, the higher frequency for Burst 3 than for the Discovery Burst implies a narrower pulse. This thus provides a natural explanation for the observed narrowing of FRB signals with increasing frequency, as exemplified in Figures \ref{fig:FRB121102-OH1612} and \ref{fig:FRB121102-OH6030} \citep{Michilli2018,Gajjar2018}.  

The significantly shorter duration for this burst ($\sim0.5$ ms) places constraints on our model since it potentially originates from the same region that produced the broader Discovery Burst at a lower frequency (see Figure \ref{fig:FRB121102-OH1612}). We therefore posit a similar length $L$ for the corresponding SR systems, while it is also reasonable to assume that the common origin for the two bursts would imply triggering from the same source (see Sec. \ref{sec:trigger}). We thus endeavoured to produce a fit for Burst 3 with a model using the same triggering pulse duration and SR sample length as the ones for the FRB 121102 Discovery Burst (i.e., $T_\mathrm{t}=0.1$ ms and $L=4.45\times10^{15}$ cm, respectively). The remaining fit parameters yielded $T_2=0.05$ ms for the dephasing time-scale (we still used $T_1=10$ s for relaxation processes) and an inverted population $nL = 1.1\times10^{14}\,\mathrm{cm}^{-2}$. The underlying density of the inverted column density for a single SR sample remains basically unchanged at $n=0.024\,\mathrm{cm}^{-3}$, which corresponds to a density of $0.12\,\mathrm{cm}^{-3}$ within $1\,\mathrm{km\,s}^{-1}$ given the wider spectral extent associated to a single SR sample (i.e., $\sim1000$ Hz). In other words, for a fixed SR sample length the Burst 3 model with the OH 6030 MHz line requires an inversion level that is approximately an order of magnitude lower than that for the Discovery Burst with the OH 1612 MHz transition. Finally, the amplitude of the triggering pulse was increased slightly to $0.75\,\mathrm{nV}\,\mathrm{m}^{-1}$, while the cross section of the SR sample has a radius of 839 km and $\sim2\times10^{30}$ inverted molecules are responsible for the SR burst.  

\subsection{Source of the triggering pulse}\label{sec:trigger}

The models above require significant amounts of molecular material, and use short coherent triggering pulses ($\sim 0.5$~ms for FRB~1102220 and 0.1~ms for FRB~121102) at levels of $\sim10^{-9}\,\mathrm{V}\,\mathrm{m}^{-1}$ or less. One obvious candidate to be considered as a source are young pulsars, which are known to emit brief coherent pulses in the frequency range we are concerned with, and which are typically born in star-forming regions near molecular clouds. We now verify whether this is a possibility by assessing the intensity of radiation needed to trigger the onset of an SR burst given the examples above. 

With the intensity of the triggering pulse given by $I_\mathrm{t} = c \epsilon_0 E_0^2/2$, it follows that a broadband triggering pulse with an electric field amplitude $E_0 = 10^{-9}\,\mathrm{V}\,\mathrm{m}^{-1}$ over a bandwidth of 1~kHz (i.e., approximately matching the corresponding parameters for Burst~3 of FRB~121102) would yield a flux density of $\sim 100$~Jy. At a distance of $\sim 100$~pc from a SR/FRB system and at a frequency of $\sim 1.4$~GHz, eight of the $\sim1800$ known radio pulsars would have a pulse-averaged flux density $>100$~Jy \citep{Manchester2005}.\footnote{ATNF Pulsar Catalogue v1.58, accessed on 2018 August 12} Of particular note as an indicative example of feasibility is PSR~J1410--6132, an energetic radio and $\gamma$-ray pulsar with a characteristic age of just $\sim$25\,000 years and at Galactic coordinates $\ell = 312.2-$ and $b = -0.09$ \citep{OBrien2008}. PSR~J1410--6132 has a flux density of $6\pm1$~mJy at 1.5~GHz at an estimated distance of $\sim15$~kpc, with a narrow pulse of full width $\sim2$~ms. At a distance of 100~pc, this would imply a flux density of $\sim135$~Jy (pulse-averaged) and $\sim3000$~Jy (peak). This pulsar is also just 20~pc from the Galactic plane, where much of the molecular material in the Milky Way is found. It is inevitable that there are similarly young radio pulsars in many other galaxies that produce narrow high-fluence pulses in the vicinity of molecular clouds. 

Alternatively, two young pulsars, B0531+21 (the Crab pulsar) and B0540--69 (the ``Crab twin'' in the Large Magellanic Cloud), have pulse-averaged fluxes that would be far below 100~Jy at 100~pc, but can produce individual ``giant pulses'' that can be extremely brief and extremely bright \citep{Johnston2003,Knight2006}.\footnote{Several recycled, or ``millisecond'' pulsars also produce giant pulses. However, such systems are extremely old, and are almost never found anywhere near dense molecular material.} For example, \citet{Hankins2003} have reported micro-structure in individual giant pulses from PSR~B0531+21 with durations of a few nanoseconds, and with peak flux densities of $>1000$~Jy at a distance of $\sim2$~kpc. 

In either case, we are left with a scenario in which a straightforward triggering source in the form of a young radio pulsar, located at some astronomical distance (hundreds of pc and more), could be responsible for the relatively small incident signal ($\sim100$~Jy) needed at the input of an SR system to cause the onset of a powerful FRB detectable over cosmological distances. Incidentally, the nature of the pulsar signal could lead to the existence of repeating FRB signals. We also note that the periodicity of such a signal may not be clear from the data because of the different delay times between the trigger and the appearance of the FRB pulses depending on the frequency (i.e., spectral lines) and the inverted column density (see \citealt{Houde2018a}). The response of the SR/FRB system could be further complicated if the pulsar's period is shorter than the time interval needed to replenish the inverted population after it has been quenched by the emission of a radiation pulse. Subsequent bursts of radiation could therefore exhibit significantly differing intensities, some being too low for detection.

It is also interesting to note that there exists a least one reported case of a background pulsar stimulating the emission of OH 1720 MHz photons from a foreground molecular cloud \citep{Weisberg2005}. Although the physical conditions in this case were not sufficient to sustain even a maser action, it is nonetheless similar in form to the scenario described above.

Incidentally, triggered SR can easily account for strongly polarized FRB signals, as seen in recent observations of FRB 121102 \citep{Michilli2018,Gajjar2018}. That is, the triggering pulse, if polarized, will couple more efficiently to a spectral transition that has similar polarization characteristics. This could then favour one transition out of a group of degenerate lines and lead to the presence of large polarization levels in the FRB. 

\subsection{Spectral bandwidth}

\begin{center}
   \begin{figure}
        \includegraphics[width=\columnwidth]{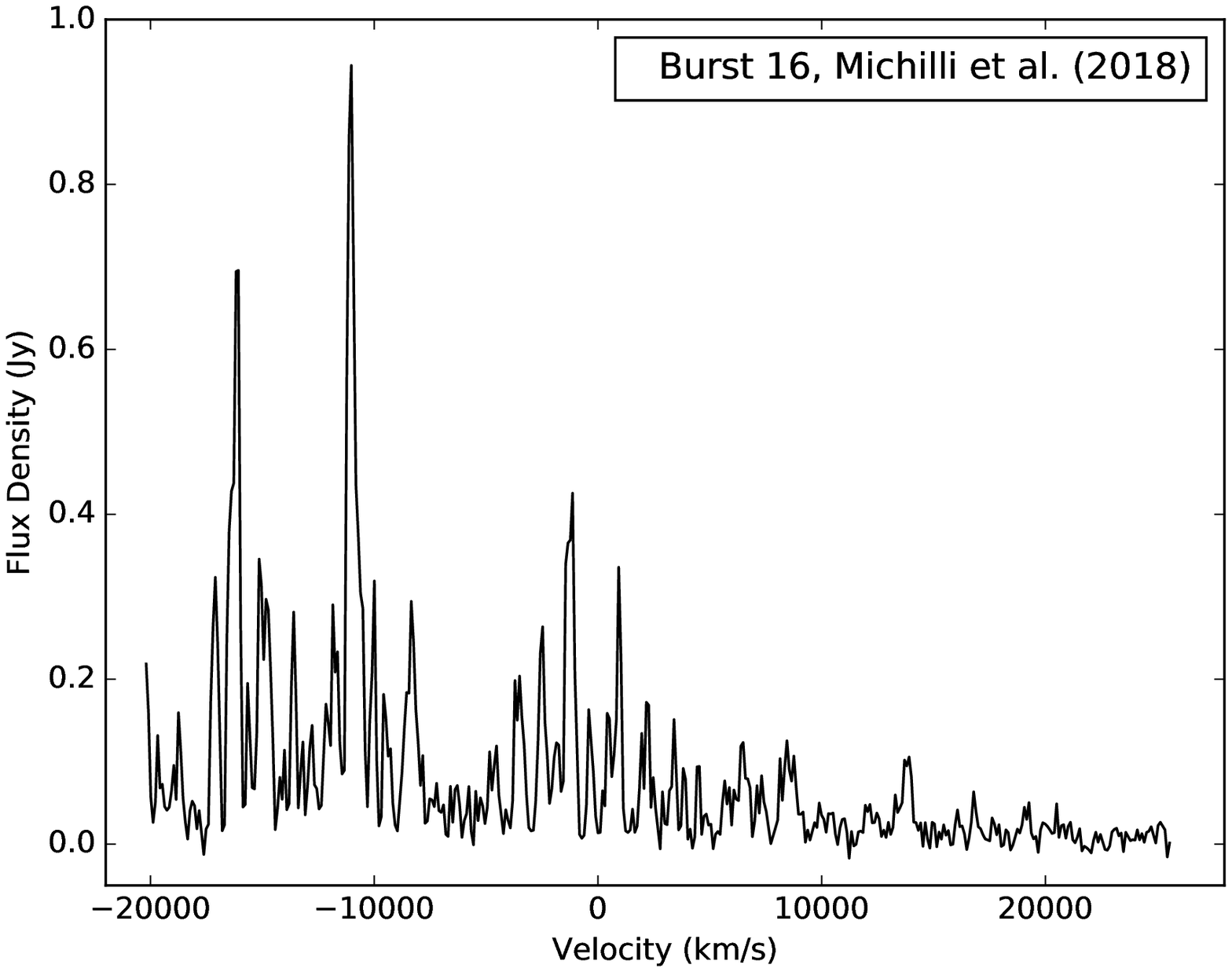}
        \includegraphics[width=\columnwidth]{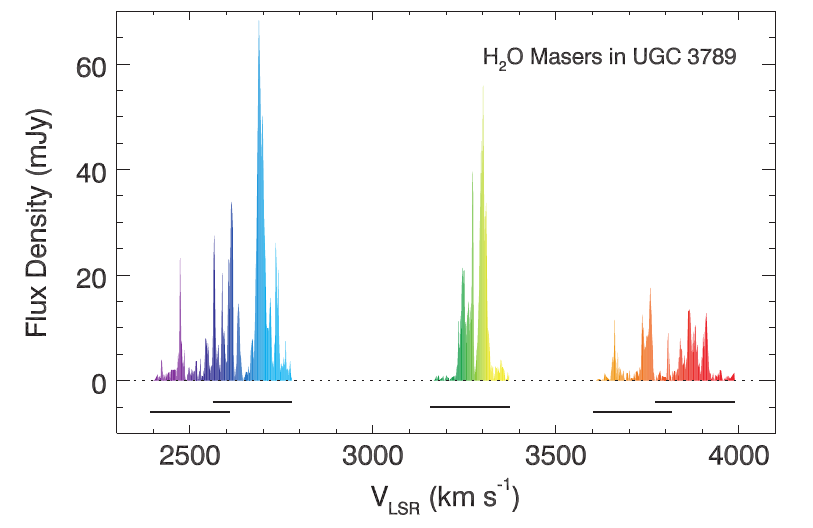}
        \caption{Top: the spectrum of Burst 16 (adapted) from \citet{Michilli2018} as a function of velocity. To do so we arbitrarily chose $\nu_0=4.55$~GHz as the centre frequency and determined the velocity with $\mathrm{v}=-c\,\Delta\nu/\nu_0$, where $\Delta\nu$ is the excursion from $\nu_0$. Bottom: the spectrum of the $\mathrm{H_2O}$ megamaser toward UGC 3789, taken from Figure 1 of \citet{Reid2009}. Although the velocity range covered by the FRB signal is significantly larger, the structure of the FRB spectrum is reminiscing of that observed for the megamaser.} 
        \label{fig:Burst16}
    \end{figure}
\end{center}

\begin{center}
   \begin{figure}
        \includegraphics[width=\columnwidth]{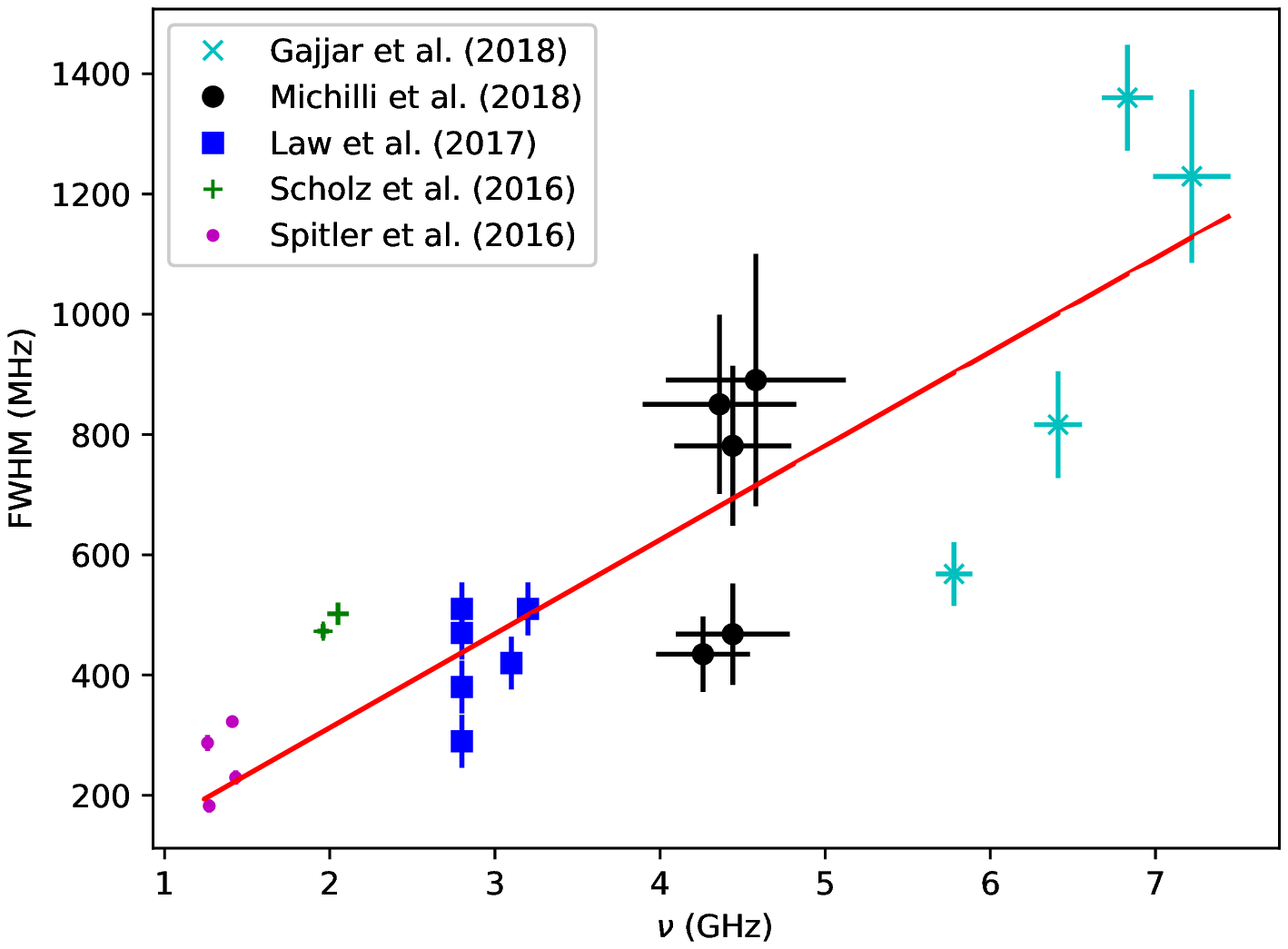}
        \caption{The bandwidth (FWHM in MHz) of FRB 121102 signals measured as a function of the frequency band ($\nu$ in GHz) at which they were measured. The data were taken from \citet{Spitler2016} (at $\sim1.4$ GHz), \citet{Scholz2016} (at $\sim2$ GHz), \citet{Law2017} (at $\sim3$ GHz), \citet{Michilli2018} (at 4--5 GHz) and \citet{Gajjar2018} (at 5--8 GHz; Bursts 11A and 11D).  In all cases the spectra were integrated over a time interval containing the FRB signal and fitted with a simple Gaussian function. The solid red line is a linear fit of the form $\mathrm{FWHM}=a\nu$, with $a=156\pm4$ MHz/GHz, as would be expected for a relation based on the (non-relativistic) Doppler effect alone.}
        \label{fig:FWHM}
    \end{figure}
\end{center}

In our SR model for FRBs, \emph{a single} SR sample is responsible for the emission of a FRB of narrow spectral extent (e.g., $\sim1$ kHz for Burst 3 of \citealt{Michilli2018} presented in Figure \ref{fig:FRB121102-OH6030}) centred at the frequency of the corresponding molecular transition. It therefore follows that the broad FRB bandwidths must be due to gas motions covering the large (mildly relativistic) velocity range needed to produce the wide spectral widths at the observed frequencies \citep{Houde2018a}. A large number of independent SR systems of small spectral width, but Doppler shifted relative to each other, then fill up the total FRB spectrum. Although the notion that the wide spectra of FRBs could result from a process based on spectral lines, which are intrinsically narrowband, could at first sight appear unlikely, there are well known cases in astrophysics where this happens. To make this point clearer, we show in the top panel of Figure \ref{fig:Burst16} the spectrum of Burst 16 of \citet{Michilli2018} displayed as a function of velocity (we arbitrarily chose $\nu_0=4.55$~GHz as the centre frequency and determined the velocity with $\mathrm{v}=-c\,\Delta\nu/\nu_0$, where $\Delta\nu$ is the excursion from $\nu_0$). Although the velocity range covered by the FRB signal is significantly larger, the structure of the spectrum is reminiscing of those observed for megamasers, which are based on an underlying physical process (stimulated emission) that is narrowband in nature. This can be asserted, for example, through a comparison with the spectrum of the $\mathrm{H_2O}$ megamaser toward UGC 3789 also shown in Figure \ref{fig:Burst16} (bottom panel; taken from Figure 1 of \citealt{Reid2009}). If this situation also holds for FRBs, then we expect their spectral bandwidths to approximately scale linearly with frequency through the Doppler effect.
We show in Figure \ref{fig:FWHM} a graph of the bandwidth (FWHM in MHz) of FRB 121102 signals measured as a function of the frequency band ($\nu$ in GHz) at which they were measured. The data were taken from \citet{Spitler2016} (at $\sim1.4$ GHz), \citet{Scholz2016} (at $\sim2$ GHz), \citet{Law2017} (at $\sim3$ GHz), \citet{Michilli2018} (at 4--5 GHz) and \citet{Gajjar2018} (at 5--8 GHz; Bursts 11A and 11D).  For all cases, except the data from \citet{Law2017} and \citet{Gajjar2018}, the spectra were integrated over a time interval containing their corresponding FRB signal and fitted with a simple Gaussian function, yielding an amplitude, a frequency centre ($\nu$) and a spectral width (FWHM). A similar analysis was already performed by the authors for the data from \citet{Law2017} and their results presented in their Table 2, which we have used in Figure \ref{fig:FWHM}. For the \citet{Gajjar2018} data a two-component Gaussian fit was required for some spectra, as the presence of significant fine structure of narrower bandwidth needed to be fitted along with the broad component we were seeking to evaluate. Moreover, whenever multiple pulses appeared at different times and frequencies in a dynamic spectrum they were treated and fitted separately (e.g., see FRB 11A in Figure 2 of \citealt{Gajjar2018}). For all data sets we limited ourselves to spectra with sufficient SNR and avoided cases where the spectral profile was severely truncated at one edge of the corresponding frequency band. The large uncertainties at the higher frequency bands are mostly due to the presence of significant fine structure in the spectra \citep{Law2017} and the fact that the assumed Gaussian form is not perfectly realized in the data.

 We also note that a linear relationship between the bandwidth and the frequency would imply that the different spectral transitions responsible for the emission across the bands are all excited by the same gas components. This is unlikely to be realized as the excitation requirements vary from line to line. For example, the groups of OH $^2\Pi_{3/2}$ ($J=3/2$) 1.7 GHz, $^2\Pi_{3/2}$ ($J=5/2$) 6 GHz and $^2\Pi_{1/2}$ ($J=7/2$) 5.5 GHz spectral lines have excitation temperatures of $\simeq0$ K, $\simeq120$ K and $\simeq617$ K, respectively. It follows that we should expect transitions that are more easily excited to span wider velocity ranges (and relative bandwidths), and vice-versa. Still, despite the amount of dispersion present in the measured bandwidths at basically all frequencies in Figure \ref{fig:FWHM}, there is a clear tendency for the FWHM to systematically increase with frequency. The solid red line in the figure is a linear fit of the form $\mathrm{FWHM}=a\nu$, with $a=156\pm4$ MHz/GHz, which would be expected for a relation based on the (non-relativistic) Doppler effect alone. We note that the assumptions underlying this linear fit, namely a Gaussian spectral shape for measuring the spectral widths and their linear scaling with frequency, are not necessary conditions for SR. They simply act as tools to better visualize and quantify the systematic increase of signal bandwidth with frequency.    
 
 This result is consistent with the prediction of our SR-based model that the bandwidth of an FRB is due to the velocity range covered by motions in the gas where the radiation is originating. We thus interpret this behaviour as consistent with our proposal that FRBs result from emission due to spectral lines.

\section{Conclusion}\label{sec:conclusion}

In this paper, we have applied a triggered SR model to FRBs and, in particular, to previously published data of FRB 110220 \citep{Thornton2013} and FRB 121102 \citep{Spitler2014,Michilli2018}. We have shown how, for example, a young pulsar located at $\sim100$ pc or more from an SR/FRB system could initiate the onset of a powerful burst of radiation detectable over cosmological distances. In the process the electric field of the coherent triggering pulse is ``amplified'' by a factor of approximately $10^8$ by the SR system. 

Our models using the OH $^2\Pi_{3/2}$ ($J=3/2$) 1612 MHz and $^2\Pi_{3/2}$ ($J=5/2$) 6030 MHz spectral lines match the light curves well and suggest the entanglement of more than $10^{30}$ initially inverted molecules over lengths of approximately 300 au for a single SR sample (the inverted column density $nL\sim10^{14}\,\mathrm{cm}^{-2}$). The FRB systems responsible for the emission of the detected pulses are expected to be composed of a very large number of such SR samples that span the observed large bandwidths and cover the expected spatial volume occupied by inverted molecular populations (e.g., as in maser- or megamaser-hosting environments). 

Furthermore, our SR model for FRB 121102 naturally accounts for the observed temporal narrowing of FRB pulses with increasing frequency \citep{Michilli2018,Gajjar2018}. We also predict an approximately linear scaling between the FRB spectral bandwidth and the frequency of observation, which we found to be consistent with the existing data. Finally, the high Faraday rotation observed in FRB 121102 could be accounted for in our SR model by the fact that it requires the presence of significant amounts of molecular material for the generation of the corresponding signals.

\section*{Acknowledgements}

We thank J.W.T. Hessels for sharing the FRB 121102 data published in \citet{Michilli2018} and \citet{Gajjar2018}, as well as for preparing them in a format we could more easily handle. We also thank M. Chamma and C. Wyenberg for their help in preparing Figure 4. M.H.'s research is funded through the Natural Sciences and Engineering Research Council of Canada Discovery Grant RGPIN-2016-04460. The Dunlap Institute is funded through an endowment established by the David Dunlap family and the University of Toronto. B.M.G. acknowledges the support of the Natural Sciences and Engineering Research Council of Canada (NSERC) through grant RGPIN-2015-05948, and of the Canada Research Chairs program. A.M. is supported by the U.S. Department of Energy (DOE) Office of Science Fusion Energy Sciences program under contract DE-SC0014264 and the Joseph P. Kearney Fellowship.




\bibliographystyle{mnras}
\bibliography{FRB-bib} 

\begin{thebibliography}{}
\makeatletter
\relax
\def\mn@urlcharsother{\let\do\@makeother \do\$\do\&\do\#\do\^\do\_\do\%\do\~}
\def\mn@doi{\begingroup\mn@urlcharsother \@ifnextchar [ {\mn@doi@}
  {\mn@doi@[]}}
\def\mn@doi@[#1]#2{\def\@tempa{#1}\ifx\@tempa\@empty \href
  {http://dx.doi.org/#2} {doi:#2}\else \href {http://dx.doi.org/#2} {#1}\fi
  \endgroup}
\def\mn@eprint#1#2{\mn@eprint@#1:#2::\@nil}
\def\mn@eprint@arXiv#1{\href {http://arxiv.org/abs/#1} {{\tt arXiv:#1}}}
\def\mn@eprint@dblp#1{\href {http://dblp.uni-trier.de/rec/bibtex/#1.xml}
  {dblp:#1}}
\def\mn@eprint@#1:#2:#3:#4\@nil{\def\@tempa {#1}\def\@tempb {#2}\def\@tempc
  {#3}\ifx \@tempc \@empty \let \@tempc \@tempb \let \@tempb \@tempa \fi \ifx
  \@tempb \@empty \def\@tempb {arXiv}\fi \@ifundefined
  {mn@eprint@\@tempb}{\@tempb:\@tempc}{\expandafter \expandafter \csname
  mn@eprint@\@tempb\endcsname \expandafter{\@tempc}}}

\bibitem[\protect\citeauthoryear{{Arecchi} \& {Courtens}}{{Arecchi} \&
  {Courtens}}{1970}]{Arecchi1970}
{Arecchi} F.~T.,  {Courtens} E.,  1970, \mn@doi [\pra]
  {10.1103/PhysRevA.2.1730}, \href
  {http://adsabs.harvard.edu/abs/1970PhRvA...2.1730A} {2, 1730}

\bibitem[\protect\citeauthoryear{Benedict et~al.}{Benedict
  et~al.}{1996}]{Benedict1996}
Benedict M.~G.,  et~al., 1996, Super-radiance: Multiatomic Coherent Emission.
IOP Publishing Ltd

\bibitem[\protect\citeauthoryear{{Brito}, {Cardoso}  \& {Pani}}{{Brito}
  et~al.}{2015}]{Brito2015}
{Brito} R.,  {Cardoso} V.,   {Pani} P.,  eds, 2015, {Superradiance}  Lecture
  Notes in Physics, Berlin Springer Verlag Vol. 906.
 (\mn@eprint {arXiv} {1501.06570})

\bibitem[\protect\citeauthoryear{Carlson, Jackson, Schawlow, Gross  \&
  Haroche}{Carlson et~al.}{1980}]{Carlson1980}
Carlson N.,  Jackson D.,  Schawlow A.,  Gross M.,   Haroche S.,  1980, Optics
  Communications, 32, 350

\bibitem[\protect\citeauthoryear{{Chatterjee} et~al.,}{{Chatterjee}
  et~al.}{2017}]{Chatterjee2017}
{Chatterjee} S.,  et~al., 2017, \mn@doi [\nat] {10.1038/nature20797}, \href
  {http://adsabs.harvard.edu/abs/2017Natur.541...58C} {541, 58}

\bibitem[\protect\citeauthoryear{{Conlon} \& {Herdeiro}}{{Conlon} \&
  {Herdeiro}}{2018}]{Conlon2018}
{Conlon} J.~P.,  {Herdeiro} C.~A.~R.,  2018, \mn@doi [Physics Letters B]
  {10.1016/j.physletb.2018.02.073}, \href
  {http://adsabs.harvard.edu/abs/2018PhLB..780..169C} {780, 169}

\bibitem[\protect\citeauthoryear{Dicke}{Dicke}{1954}]{Dicke1954}
Dicke R.~H.,  1954, \mn@doi [Phys. Rev.] {10.1103/PhysRev.93.99}, \href
  {http://adsabs.harvard.edu/abs/1954PhRv...93...99D} {93, 99}

\bibitem[\protect\citeauthoryear{Dicke}{Dicke}{1964}]{Dicke1964}
Dicke R.~H.,  1964, Quantum electron., 1, 35

\bibitem[\protect\citeauthoryear{{Fialkov} \& {Loeb}}{{Fialkov} \&
  {Loeb}}{2017}]{Fialkov2017a}
{Fialkov} A.,  {Loeb} A.,  2017, \mn@doi [\apjl] {10.3847/2041-8213/aa8905},
  \href {http://adsabs.harvard.edu/abs/2017ApJ...846L..27F} {846, L27}

\bibitem[\protect\citeauthoryear{{Fialkov}, {Loeb}  \& {Lorimer}}{{Fialkov}
  et~al.}{2017}]{Fialkov2017b}
{Fialkov} A.,  {Loeb} A.,   {Lorimer} D.~R.,  2017, preprint, \href
  {http://adsabs.harvard.edu/abs/2017arXiv171104396F} {} (\mn@eprint {arXiv}
  {1711.04396})

\bibitem[\protect\citeauthoryear{{Gajjar} et~al.,}{{Gajjar}
  et~al.}{2018}]{Gajjar2018}
{Gajjar} V.,  et~al., 2018, preprint, \href
  {http://adsabs.harvard.edu/abs/2018arXiv180404101G} {} (\mn@eprint {arXiv}
  {1804.04101})

\bibitem[\protect\citeauthoryear{Gray}{Gray}{2012}]{Gray2012}
Gray M.,  2012, Maser Sources in Astrophysics.
Cambridge University Press

\bibitem[\protect\citeauthoryear{Gross \& Haroche}{Gross \&
  Haroche}{1982}]{Gross1982}
Gross M.,  Haroche S.,  1982, \physrep, 93, 301

\bibitem[\protect\citeauthoryear{Gross, Fabre, Pillet  \& Haroche}{Gross
  et~al.}{1976}]{Gross1976}
Gross M.,  Fabre C.,  Pillet P.,   Haroche S.,  1976, \prl, 36, 1035

\bibitem[\protect\citeauthoryear{{Hankins}, {Kern}, {Weatherall}  \&
  {Eilek}}{{Hankins} et~al.}{2003}]{Hankins2003}
{Hankins} T.~H.,  {Kern} J.~S.,  {Weatherall} J.~C.,   {Eilek} J.~A.,  2003,
  \mn@doi [\nat] {10.1038/nature01477}, \href
  {http://adsabs.harvard.edu/abs/2003Natur.422..141H} {422, 141}

\bibitem[\protect\citeauthoryear{{Haroche}, {Fabre}, {Gross}  \&
  {Pillet}}{{Haroche} et~al.}{1977}]{Haroche1977}
{Haroche} S.,  {Fabre} C.,  {Gross} M.,   {Pillet} P.,  1977, in {Marrus} R.,
  {Prior} M.,   {Shugart} H.,  eds, Atomic Physics 5. p.~179

\bibitem[\protect\citeauthoryear{Houde \& Rajabi}{Houde \&
  Rajabi}{2018}]{Houde2018b}
Houde M.,  Rajabi F.,  2018, Journal of Physics Communications, 2, 075015

\bibitem[\protect\citeauthoryear{Houde, Mathews  \& Rajabi}{Houde
  et~al.}{2018}]{Houde2018a}
Houde M.,  Mathews A.,   Rajabi F.,  2018, \mn@doi [\mnras]
  {10.1093/mnras/stx3205}, \href
  {http://adsabs.harvard.edu/abs/2018MNRAS.475..514H} {475, 514}

\bibitem[\protect\citeauthoryear{{Johnston} \& {Romani}}{{Johnston} \&
  {Romani}}{2003}]{Johnston2003}
{Johnston} S.,  {Romani} R.~W.,  2003, \mn@doi [\apjl] {10.1086/376826}, \href
  {http://adsabs.harvard.edu/abs/2003ApJ...590L..95J} {590, L95}

\bibitem[\protect\citeauthoryear{{Keane} et~al.,}{{Keane}
  et~al.}{2016}]{Keane2016}
{Keane} E.~F.,  et~al., 2016, \mn@doi [\nat] {10.1038/nature17140}, \href
  {http://adsabs.harvard.edu/abs/2016Natur.530..453K} {530, 453}

\bibitem[\protect\citeauthoryear{{Knight}}{{Knight}}{2006}]{Knight2006}
{Knight} H.~S.,  2006, Chinese Journal of Astronomy and Astrophysics
  Supplement, \href {http://adsabs.harvard.edu/abs/2006ChJAS...6b..41K} {6, 41}

\bibitem[\protect\citeauthoryear{{Law} et~al.,}{{Law} et~al.}{2017}]{Law2017}
{Law} C.~J.,  et~al., 2017, \mn@doi [\apj] {10.3847/1538-4357/aa9700}, \href
  {http://adsabs.harvard.edu/abs/2017ApJ...850...76L} {850, 76}

\bibitem[\protect\citeauthoryear{{Lorimer}, {Bailes}, {McLaughlin}, {Narkevic}
  \& {Crawford}}{{Lorimer} et~al.}{2007}]{Lorimer2007}
{Lorimer} D.~R.,  {Bailes} M.,  {McLaughlin} M.~A.,  {Narkevic} D.~J.,
  {Crawford} F.,  2007, \mn@doi [\sci] {10.1126/science.1147532}, \href
  {http://adsabs.harvard.edu/abs/2007Sci...318..777L} {318, 777}

\bibitem[\protect\citeauthoryear{{Manchester}, {Hobbs}, {Teoh}  \&
  {Hobbs}}{{Manchester} et~al.}{2005}]{Manchester2005}
{Manchester} R.~N.,  {Hobbs} G.~B.,  {Teoh} A.,   {Hobbs} M.,  2005, \mn@doi
  [\aj] {10.1086/428488}, \href
  {http://adsabs.harvard.edu/abs/2005AJ....129.1993M} {129, 1993}

\bibitem[\protect\citeauthoryear{{Marcote} et~al.,}{{Marcote}
  et~al.}{2017}]{Marcote2017}
{Marcote} B.,  et~al., 2017, \mn@doi [\apjl] {10.3847/2041-8213/834/2/L8},
  \href {http://adsabs.harvard.edu/abs/2017ApJ...834L...8M} {834, L8}

\bibitem[\protect\citeauthoryear{Mathews}{Mathews}{2017}]{Mathews2017}
Mathews A.,  2017, The Role of Superradiance in Cosmic Fast Radio Bursts,
  Honours thesis, The University of Western Ontario

\bibitem[\protect\citeauthoryear{{Michilli} et~al.,}{{Michilli}
  et~al.}{2018}]{Michilli2018}
{Michilli} D.,  et~al., 2018, \mn@doi [\nat] {10.1038/nature25149}, \href
  {http://adsabs.harvard.edu/abs/2018Natur.553..182M} {553, 182}

\bibitem[\protect\citeauthoryear{{O'Brien} et~al.,}{{O'Brien}
  et~al.}{2008}]{OBrien2008}
{O'Brien} J.~T.,  et~al., 2008, \mn@doi [\mnras]
  {10.1111/j.1745-3933.2008.00481.x}, \href
  {http://adsabs.harvard.edu/abs/2008MNRAS.388L...1O} {388, L1}

\bibitem[\protect\citeauthoryear{{Petroff} et~al.,}{{Petroff}
  et~al.}{2016}]{Petroff2016}
{Petroff} E.,  et~al., 2016, \mn@doi [Proc. Astron. Soc. Au.]
  {10.1017/pasa.2016.35}, \href
  {http://adsabs.harvard.edu/abs/2016PASA...33...45P} {33, e045}

\bibitem[\protect\citeauthoryear{Rajabi}{Rajabi}{2016}]{Rajabi2016Thesis}
Rajabi F.,  2016, PhD thesis, The University of Western Ontario

\bibitem[\protect\citeauthoryear{Rajabi \& Houde}{Rajabi \&
  Houde}{2016a}]{Rajabi2016A}
Rajabi F.,  Houde M.,  2016a, \apj, 826, 216

\bibitem[\protect\citeauthoryear{Rajabi \& Houde}{Rajabi \&
  Houde}{2016b}]{Rajabi2016B}
Rajabi F.,  Houde M.,  2016b, \apj, 828, 57

\bibitem[\protect\citeauthoryear{{Rajabi} \& {Houde}}{{Rajabi} \&
  {Houde}}{2017}]{Rajabi2017}
{Rajabi} F.,  {Houde} M.,  2017, \mn@doi [Science Advances]
  {10.1126/sciadv.1601858}, \href
  {http://adsabs.harvard.edu/abs/2017SciA....3E1858R} {3, e1601858}

\bibitem[\protect\citeauthoryear{{Ravi} et~al.,}{{Ravi}
  et~al.}{2016}]{Ravi2016}
{Ravi} V.,  et~al., 2016, \mn@doi [\sci] {10.1126/science.aaf6807}, \href
  {http://adsabs.harvard.edu/abs/2016Sci...354.1249R} {354, 1249}

\bibitem[\protect\citeauthoryear{{Reid}, {Braatz}, {Condon}, {Greenhill},
  {Henkel}  \& {Lo}}{{Reid} et~al.}{2009}]{Reid2009}
{Reid} M.~J.,  {Braatz} J.~A.,  {Condon} J.~J.,  {Greenhill} L.~J.,  {Henkel}
  C.,   {Lo} K.~Y.,  2009, \mn@doi [\apj] {10.1088/0004-637X/695/1/287}, \href
  {http://adsabs.harvard.edu/abs/2009ApJ...695..287R} {695, 287}

\bibitem[\protect\citeauthoryear{{Rosa} \& {Kephart}}{{Rosa} \&
  {Kephart}}{2018}]{Rosa2018}
{Rosa} J.~G.,  {Kephart} T.~W.,  2018, \mn@doi [Physical Review Letters]
  {10.1103/PhysRevLett.120.231102}, \href
  {http://adsabs.harvard.edu/abs/2018PhRvL.120w1102R} {120, 231102}

\bibitem[\protect\citeauthoryear{{Scholz} et~al.,}{{Scholz}
  et~al.}{2016}]{Scholz2016}
{Scholz} P.,  et~al., 2016, \mn@doi [\apj] {10.3847/1538-4357/833/2/177}, \href
  {http://adsabs.harvard.edu/abs/2016ApJ...833..177S} {833, 177}

\bibitem[\protect\citeauthoryear{{Spitler} et~al.,}{{Spitler}
  et~al.}{2014}]{Spitler2014}
{Spitler} L.~G.,  et~al., 2014, \mn@doi [\apj] {10.1088/0004-637X/790/2/101},
  \href {http://adsabs.harvard.edu/abs/2014ApJ...790..101S} {790, 101}

\bibitem[\protect\citeauthoryear{{Spitler} et~al.,}{{Spitler}
  et~al.}{2016}]{Spitler2016}
{Spitler} L.~G.,  et~al., 2016, \mn@doi [\nat] {10.1038/nature17168}, \href
  {http://adsabs.harvard.edu/abs/2016Natur.531..202S} {531, 202}

\bibitem[\protect\citeauthoryear{{Tendulkar} et~al.,}{{Tendulkar}
  et~al.}{2017}]{Tendulkar2017}
{Tendulkar} S.~P.,  et~al., 2017, \mn@doi [\apjl] {10.3847/2041-8213/834/2/L7},
  \href {http://adsabs.harvard.edu/abs/2017ApJ...834L...7T} {834, L7}

\bibitem[\protect\citeauthoryear{{Thornton} et~al.,}{{Thornton}
  et~al.}{2013}]{Thornton2013}
{Thornton} D.,  et~al., 2013, \mn@doi [\sci] {10.1126/science.1236789}, \href
  {http://adsabs.harvard.edu/abs/2013Sci...341...53T} {341, 53}

\bibitem[\protect\citeauthoryear{{Weisberg}, {Johnston}, {Koribalski}  \&
  {Stanimirovi{\'c}}}{{Weisberg} et~al.}{2005}]{Weisberg2005}
{Weisberg} J.~M.,  {Johnston} S.,  {Koribalski} B.,   {Stanimirovi{\'c}} S.,
  2005, \mn@doi [Science] {10.1126/science.1112494}, \href
  {http://adsabs.harvard.edu/abs/2005Sci...309..106W} {309, 106}

\makeatother
\end{thebibliography}






\bsp	
\label{lastpage}
\end{document}